\documentclass{pasj00}

\begin{document}
\SetRunningHead{T. Shimizu, K. Masai, \& K. Koyama}{Non-thermal emission from SNR with CSM}

\title{Non-Thermal Radio and Gamma-Ray Emission from a Supernova Remnant by the Blast Wave Breaking Out of the Circumstellar Matter}

\author{Takafumi \textsc{Shimizu}\altaffilmark{1}, Kuniaki \textsc{Masai}\altaffilmark{1}, and Katsuji \textsc{Koyama}\altaffilmark{2,3}
}
\altaffiltext{1}{Department of Physics, Tokyo Metropolitan University, 1-1 Minami-Ohsawa, Hachioji, Tokyo 192-0397}
\email{t-shimizu@phys.se.tmu.ac.jp}
\altaffiltext{2}{Department of Physics, Graduate school of Science, Kyoto University, Oiwake-cho, Kitashirakawa, Kyoto 606-8502}
\altaffiltext{3}{Department of Earth and Space Science, Graduate School of Science, Osaka University, 1-1 Machikaneyama, Toyonaka, Osaka 560-0043}

%


\maketitle

\begin{abstract}

We calculate synchrotron radio emission and $\gamma$-ray emission due to bremsstrahlung, inverse-Compton scattering and $\pi^0$-decay from the remnant of supernova which exploded in the circumstellar matter (CSM) formed by the progenitor's stellar wind. This sort of situation is a possible origin of mixed-morphology supernova remnants (SNRs) like W49B, which exhibit recombination-radiation spectra in X-ray emission.  We assume that the CSM of $1.5~M_{\odot}$ exists at 0.07--3~pc away from the supernova in the interstellar medium (ISM) of density 0.016~cm$^{-3}$.  When the blast wave breaks out of the CSM into the ISM, its velocity rapidly increases and hence particle acceleration is enhanced.  The maximum energy of protons reaches $\sim 1300$~TeV just after the break-out with $\sim 0.5$\% of the explosion energy. We consider the non-thermal emission from the blast-shocked ISM shell after the break-out. 
Synchrotron radio flux at 1~GHz is tens Jy, comparable to the observed from mixed-morphology SNRs. 
Because of low density, the $\gamma$-ray luminosity is dominated by inverse-Compton scattering, which is higher than the $\pi^0$-decay luminosity by an order of magnitude. The total $\gamma$-ray luminosity including bremsstrahlung is of the order of $10^{33}$~erg\,s$^{-1}$ lower than the typical value $10^{35}$--$10^{36}$~erg\,s$^{-1}$ observed from mixed-morphology SNRs. However, if, e.g., $\sim10$\% of accelerated protons interact with some matter of density $\sim100$~cm$^{-3}$, $\pi^0$-decay $\gamma$-ray luminosity would be enhanced to be comparable with the observed value. 

\end{abstract}

\section{Introduction}


About 70\% of Galactic supernova remnants (SNRs) exhibit shell-like morphology in radio wavelength \citep{Green09}. The radio shells are due to synchrotron emission by non-thermal electrons of order of GeV. Many of these SNRs also exhibit shell-like morphology in X-rays, due to bremsstrahlung and line emission by thermal electrons of order of keV. The electrons are heated and/or accelerated by interstellar shocks of SNRs. On the other hand, $\sim10$\% of the SNRs that have the radio shells exhibit center-filled thermal X-rays, and are called mixed-morphology (MM) SNRs (e.g., \cite{Rho98}). 
Recently, {\it Suzaku} revealed recombination radiation, which are hardly expected for typical shell-like SNRs, in X-ray spectra of six MM SNRs so far observed: IC443 \citep{Yamaguchi09}, W49B \citep{Ozawa09}, G359.1-0.5 \citep{Ohnishi11}, W28 \citep{Sawada12}, W44 \citep{Uchida12}, and G346.6-0.2 \citep{Yamauchi12}. 

MM SNRs are characteristic also in $\gamma$-ray emission. {\it Fermi} detected GeV $\gamma$-rays from SNRs: while the luminosities in the 1--100 GeV band of shell-like SNRs are $10^{33}$--$10^{35}$ erg s$^{-1}$ (\cite{Abdo10b}, \cite{Castro10}, \cite{Katagiri11}, \cite{Abdo11}, \cite{Giordano11}, \cite{Tanaka11}), the luminosities of MM SNRs are distinctively higher, $10^{35}$--$10^{36}$ erg s$^{-1}$ (\cite{Abdo09}, \cite{Abdo10a}, \cite{Abdo10c}, \cite{Abdo10d}, \cite{Abdo10e}, \cite{Castro10}).  This may imply that such intense $\gamma$-rays come from the dense matter around MM SNRs.  Actually, for these MM SNRs, interaction with molecular clouds is suggested by 1720~MHz OH maser (\cite{Frail94}, \cite{Yusef-Zadeh95}, \cite{Frail96}, \cite{Green97}, \cite{Claussen97}, \cite{Hewitt09}) and/or near-infrared observations (\cite{Keohane07}). The $\gamma$-ray spectra of observed SNRs are not always single power-laws but exhibit a break at an energy of $\sim$1--5~GeV above which their slopes are steepen.

X-ray characteristics different than shell-like SNRs and possible interaction with molecular clouds may suggest that MM SNRs are remnants of core-collapse explosion of massive stars surrounded by H\,II regions, stellar wind matter, and molecular clouds, as in star-forming regions. Recombination radiation in X-rays are predicted by \citet{Itoh89} for the remnant of a supernova  surrounded by its progenitor's stellar wind matter. They show the X-ray spectra due to rarefaction (adiabatic cooling) caused when the blast wave breaks out of the wind matter to expand rapidly into the ambient interstellar medium.  Recently, \citet{Shimizu12}, hereafter Paper I, extend this work to non-spherically-symmetric stellar wind matter, and find that recombination X-rays exhibit center-filled morphology like MM SNRs with various shapes depending on the viewing direction.  They also suggest that synchrotron radio shell is located outside, surrounding the X-ray emitting region. 

If rarefaction caused by the shock break-out is the origin of recombination X-rays found in MM SNRs, it is naturally of our interest whether the SNR model can explain observed radio and $\gamma$-ray emissions as well.  Hence, in the present paper, we investigate non-thermal particles, which can be accelerated by the shock of the SNR model in Paper I, and emission thereby from the blast-shocked ISM shell.  In the following section, we describe the SNR model in Paper I and the particle acceleration.  In Section 3 we show calculations of non-thermal radio flux and $\gamma$-ray luminosity, and discuss the results in Section 4.

\section{Model}

\subsection{Supernova remnant}

We consider model B2 of Paper I for a model of MM SNR. The model describes evolution that the initially spherically-symmetric ejecta interact with anisotropic circumstellar matter (CSM). Outside the CSM, a uniform interstellar medium (ISM) of density $0.016$~cm$^{-3}$ is assumed. Such low density can be possible for an H\,II region formed by the progenitor. The ejecta have an initial kinetic energy of $2\times10^{51}$~erg and a mass of $10~M_{\odot}$, and therefore the initial velocity of the blast wave is $8.5\times10^3$~km s$^{-1}$. For comparison, we also calculate the evolution of SNR without CSM, which expands directly into the uniform ISM of the same density. 

The CSM is composed of the stellar wind matter blown by the progenitor in its pre-supernova phase. If the progenitor rotated rapidly, the stellar wind may have an anisotropic density distribution. We assume that the CSM is concentrated in the equatorial plane. The density on the equatorial plane is 4 times higher than that in the polar direction at the same radius. The mass-loss rate is $5\times10^{-5}~M_{\odot}$ yr$^{-1}$ at a wind velocity of $100$~km s$^{-1}$. The inner and outer radii of the CSM are 0.07~pc and 3~pc in the equatorial direction. These radii imply that the wind activity lasts $3\times10^{4}$ years and then ceases $6\times10^2$ years before the explosion. The mass of the CSM is $1.5~M_{\odot}$, which is obtained from the period of the wind activity and the mass-loss rate.

The wind parameters can be possible for B[e] supergiants or luminous blue variables (LBVs). For example, radio observations suggest that W9, which is a B[e] supergiant in Westerlund 1, has the wind velocity of $\sim10^2$~km~s$^{-1}$ and the mass-loss rate of $\sim10^{-4}~M_{\odot}$~yr$^{-1}$ \citep{Dougherty10}. LBVs have typically the wind velocity of $\sim10^2~$km~s$^{-1}$ and the mass-loss rate of $\sim10^{-5}-10^{-4}~M_{\odot}$~yr$^{-1}$ \citep{Humphreys94}. Although LBVs were considered not to explode as supernova, recent observation shows that a progenitor of SN 2005gl is a LBV \citep{Gal-Yam07}. Also, observations of type IIn supernovae suggest the interaction between ejecta and dense CSM, which have the wind velocity of $\sim10^2-10^3$~km~s$^{-1}$ and the mass-loss rate of $\sim10^{-4}-1~M_{\odot}$~yr$^{-1}$ (e.g., \cite{Kiewe12}).

\begin{figure}[htpb]
  \begin{center}
    \FigureFile(60mm,60mm){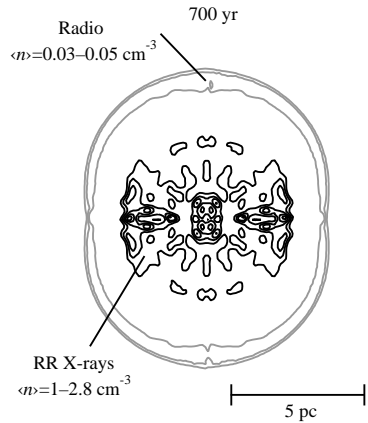}
  \end{center}
  \caption{Contours of the number density of shocked matter averaged over line of sight in the equatorial plane of model B2 at 700~yr. Black lines represent those for reverse-shocked ejecta. Gray lines represent those for the shocked ISM. ``RR X-rays" means recombination-radiation X-rays. }
  \label{fig1}
\end{figure}

The blast wave breaks out of the CSM at $\sim450$~yr in the equatorial direction, and then is rapidly accelerated. Rarefaction wave propagates inward from CSM-ISM contact interface. This causes rapid adiabatic expansion and thus cooling of the once-shocked CSM and ejecta, and results in recombination-radiation X-rays. 
Since the ISM is rarefied enough to make a density difference of factor $\sim10$ at the CSM-ISM interface, rarefaction occurs for the mass loss rate $5\times10^{-5}~M_{\odot}$ yr$^{-1}$ assumed here (cf. \cite{Moriya12}). 
After that, blast wave propagates to form a shocked shell into the ISM, while the second reverse-shock propagates inward.  Figure \ref{fig1} shows a snap shot, density map of the shocked matter averaged over the line of sight, of the model SNR. 
The inner black lines represent the reverse-shocked ejecta which is bright in thermal X-rays, and the outer gray lines represent the blast-shocked ISM which is faint in X-rays but can be bright in radio.
The late time evolution of the blast wave approaches that of the model without CSM, as shown by \citet{Itoh89}.

\begin{figure}[htpb]
  \begin{center}
    \FigureFile(80mm,90mm){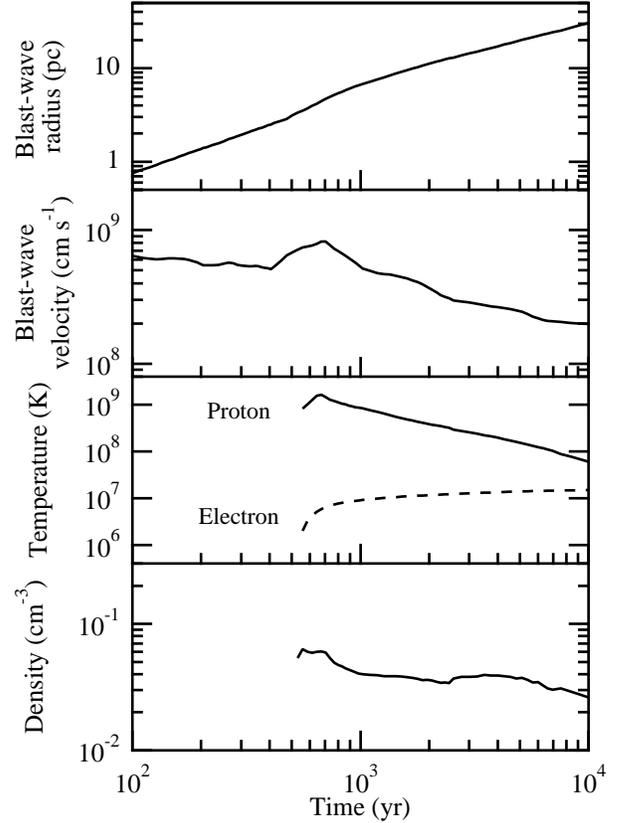}
  \end{center}
  \caption{From top to bottom, evolutions of the blast-wave radius, the blast-wave velocity, the proton and electron temperatures averaged over the blast-shocked ISM in the equatorial direction, and the number density averaged over the blast-shocked ISM in the equatorial direction, as functions of the elapsed time after explosion.}  
  \label{fig2}
\end{figure}

Figure \ref{fig2} shows the blast-wave radius $R_{\mathrm{b}}$, velocity $V_{\mathrm{s}}$, mean temperatures $\langle T_{\mathrm{p}} \rangle$ of protons and $\langle T_{\mathrm{e}} \rangle$ of electrons, and mean number density $\langle n \rangle$ of the blast-shocked ISM.  The proton and electron temperatures are calculated from the temperatures $kT_{\mathrm{p,e}} \propto m_{\mathrm{p,e}}V_{\mathrm{s}}^2$ at the shock front, assuming the energy transport from protons to electrons through Coulomb collisions in the post-shock region (\cite{Masai94}; see also Paper I), where $m_{\mathrm{p}}$ and $m_{\mathrm{e}}$ are the mass of proton and electron, respectively. 

\subsection{Particle acceleration}

We consider diffusive shock acceleration by the blast wave: part of thermal particles of the shocked ISM are scattered across the shock by magnetic inhomogeneities and gain momentum.  The energy spectrum of accelerated particles is expressed in a form \citep{Bell78}
\begin{equation}
N(E) = K(E+mc^2)(E^2+2mc^2E)^{-(\mu+1)/2}
\end{equation}
with 
\begin{equation}
K=\xi \langle n\rangle(\mu-1)(E_{\mathrm{inj}}^2+2mc^2E_{\mathrm{inj}})^{(\mu-1)/2}\nonumber
\end{equation}
for $E_{\mathrm{inj}}\leq E\leq E_{\mathrm{max}}$.  Here, $E=(\gamma-1)mc^2$ is the kinetic energy of the particle of mass $m$ with the Lorentz factor $\gamma$, $E_{\mathrm{inj}}$ is the injection kinetic energy, $E_{\mathrm{max}}$ is the maximum kinetic energy of accelerated particles, and $\xi$ is the injection efficiency.  The injection efficiency is defined as the ration of the number density of accelerated to thermal particles. 

The diffusive shock acceleration results in a single power-law energy spectrum, as described above. On the other hand, $\gamma$-ray observations of SNRs suggest that the energy spectrum of particles is not simply a single power-law, but, for instance, {\it Fermi}-observed SNRs show a break at an energy of $\sim$1--5~GeV, as mentioned in Section 1. 
Since cooling time at the break energy is much longer than the age, this break may reflect acceleration processes. 
Therefore, for the energy spectrum of accelerated particles we assume a broken power-law
\begin{equation}
N(E)=
\left\{\begin{array}{ll}
KE^{-\mu},& \mathrm{for~} E < E_{\mathrm{b}},\\
KE_{\mathrm{b}}^{-\mu+\mu_2}E^{-\mu_2},& \mathrm{for~} E \geq E_{\mathrm{b}},\\
\end{array}
\right.
\end{equation}
where $E_{\mathrm{b}}$ is the break energy of 10~GeV, taken so as to make a GeV break in the $\gamma$-ray spectrum. The spectral indexes are assumed to be $\mu=2$ and $\mu_2=2.3$, which is a medium value of spectral index of cosmic-ray sources (e.g., \cite{Putze11}).

For the injection energy $E_{\mathrm{inj}}$ and efficiency $\xi$, we consider that particles in the high energy tail of the thermal distribution are injected into acceleration process, as
\begin{equation}
E_{\mathrm{inj-p,e}}=\lambda_{\mathrm{p,e}} kT_{\mathrm{p,e}}.
\end{equation}
with a constant $\lambda$, where the characters ``p" and ``e" mean proton and electron, respectively. Then a relation between $\xi$ and $\lambda$ is given by
\begin{eqnarray}
\xi_{\mathrm{p,e}}&\equiv&\frac{\int_{E_{\mathrm{inj-p,e}}}^{\infty} f_{\mathrm{p,e}}(E)dE}{\int_{0}^{\infty} f_{\mathrm{p,e}}(E)dE}\nonumber\\
&=&1-\mathrm{erf}({\lambda_{\mathrm{p,e}}}^{1/2})+\frac{2}{\pi^{1/2}}{\lambda_{\mathrm{p,e}}}^{1/2}e^{-\lambda_{\mathrm{p,e}}},
\end{eqnarray}
where $f(E)$ is the Maxwellian distribution function, and erf is the error function.

We determine $\xi_{\mathrm{p}}$ for the pressure of accelerated protons to be equal to 10\% of the ram pressure of ISM enters in the blast wave. The pressure of accelerated particles is given by
\begin{eqnarray}
&&P_{\mathrm{CR}}=\frac{1}{3}\int_{p_{\mathrm{inj}}}^{p_{\mathrm{max}}} N'(p)pvdp\nonumber\\
&&\simeq\frac{1}{3}\xi \langle n\rangle cp_{\mathrm{inj}}\left[
\ln\left(\frac{2p_{\mathrm{b}}}{mc}\right)+\frac{1}{\mu_2-2}\left(\frac{p_{\mathrm{b}}c}{E_{\mathrm{b}}}\right)^{-\mu_2+2}
\right],
\label{equation7}
\end{eqnarray}
where we use $\mu=2$ in the last expression. Here, $v$ is the particle velocity, $p$ is the momentum of a particle, $p_{\mathrm{inj}}=(2mE_{\mathrm{inj}})^{1/2}$ is the injection momentum, $p_{\mathrm{max}}=E_{\mathrm{max}}/c$ is the maximum momentum, $p_{\mathrm{b}}$ is the break momentum, and $N'(p)dp=N(E)dE$ . In the last expression in equation (\ref{equation7}), $p_{\mathrm{inj}}\ll mc$ and $p_{\mathrm{max}} \gg p_{\mathrm{b}} > mc$ are considered. The injection efficiency of protons is roughly proportional to the blast-wave velocity $V_{\mathrm{s}}$ because $\xi_{\mathrm{p}}\propto V_{\mathrm{s}}^2/p_{\mathrm{inj-p}}\propto V_{\mathrm{s}}^2/T_{\mathrm{p}}^{1/2}\propto V_{\mathrm{s}}$. The injection efficiency of protons reaches the maximum $\sim2\times10^{-4}$ at $\sim530$~yr, and then decreases to $5\times10^{-5}$ at $\sim10000$~yr. We determine $\xi_{\mathrm{e}}$ for the pressure $P_{\mathrm{CR-e}}$ of accelerated electrons not to exceed the pressure $P_{\mathrm{CR-p}}$ of accelerated protons. The ratio of the pressure of accelerated electrons to protons is
\begin{equation}
\frac{P_{\mathrm{CR-e}}}{P_{\mathrm{CR-p}}}\simeq0.05\frac{\xi_{\mathrm{e}}}{\xi_{\mathrm{p}}}\left(\frac{E_{\mathrm{inj-e}}}{E_{\mathrm{inj-p}}}\right)^{1/2}.
\end{equation}
If the injection energy of electrons is the same as protons, $\xi_{\mathrm{e}} \lesssim20~\xi_{\mathrm{p}}$ follows. In the following, we express $\xi_{\mathrm{e}}$ in unit of $\xi_{\mathrm{p}}$. 

The maximum energy $E_{\mathrm{max}}$ is determined by the time-scales of energy gain and loss.  Adiabatic loss due to SNR expansion is negligible through the age concerned here. The dominant loss process is synchrotron radiation and inverse-Compton scattering for electrons.  Assuming that 1) mean free path of a particle is its gyration radius (Bohm limit), 2) shock compression ratio is 4, and 3) accelerated particles are relativistic ($\gamma\gg1$), we estimate the time-scales of acceleration and radiation loss as
\begin{equation}
t_{\mathrm{acc}}\simeq\frac{32\gamma mc^3}{3eBV_{\mathrm{s}}^2}
\end{equation}
and
\begin{equation}
t_{\mathrm{loss(electron)}}\simeq\frac{6\pi m_{\mathrm{e}}c}{\gamma\sigma_{\mathrm{T}}(B^2+8\pi U_{\mathrm{CMB}})},
\end{equation}
respectively, where $\sigma_{\mathrm{T}}$ is the Thomson scattering cross section, $e$ is the elementary electric charge, $B$ is the strength of the magnetic field in the shock downstream, assumed to be 4 times the strength in the upstream, and $U_{\mathrm{CMB}}$ is the energy density of cosmic microwave background.
In SNRs, the magnetic field strength can be stronger than the average interstellar value by magnetic amplification mechanisms, as suggested by X-ray variability of RX J1713.7-3946 \citep{Uchiyama07}. Using the equation (14) of \citet{Bell01}, we calculate the field strength in the SNR evolution. 
The magnetic amplification may cause the non-linear feedback from accelerated particles to the shock structure. However, such feedback is small when the injection efficiency is lower than $\sim10^{-4}$ (e.g., \cite{Ferrand10}), which is marginally attained after the blast-wave break-out concerned here.

\begin{figure}[htpb]
  \begin{center}
    \FigureFile(80mm,60mm){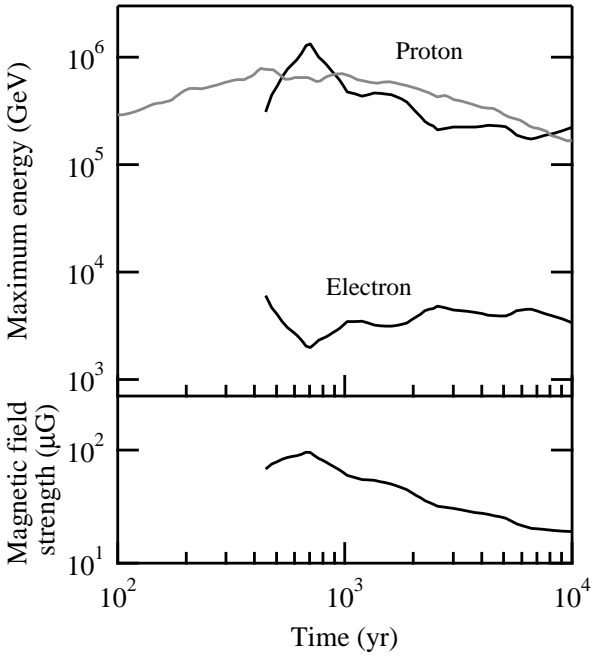}
  \end{center}
      \caption{Maximum energy that the accelerated protons and electrons can reaches at a given time ({\it upper}) and the magnetic fields strength in the shock downstream ({\it lower}), as functions of the elapsed time after explosion. Gray line represents the maximum energy of the protons in the SNR evolution without CSM.}  
  \label{fig3}
\end{figure}

In Figure \ref{fig3} we show the time evolution of $E_{\mathrm{max}}$ and $B$. For protons $E_{\mathrm{max}}$ is determined by $t_{\rm acc} \sim t_{\rm age}$ and reaches $\sim 1300$~TeV at $\sim 700$~yr, while $\sim800$~TeV at $\sim430$~yr in the case without CSM. For electrons $E_{\mathrm{max}}$ is determined by $t_{\mathrm{acc}} \sim t_{\mathrm{loss(electron)}}$, and its maximum is about 10~TeV at the moment of the break-out.  At $\sim 700$~yr just after the break-out, $E_{\mathrm{max}}$ takes its maximum/minimum for protons/electrons because of rapid increase of the shock velocity and magnetic field.  For the explosion energy $2 \times 10^{51}$~erg assumed, the total energy of accelerated protons is $1\times10^{49}$~erg at $\sim700$~yr and $\sim2\times10^{50}$~erg at $\sim10000$~yr. 

\section{Non-thermal radiation}

\begin{figure*}[htpb]
  \begin{center}
    \FigureFile(140mm,70mm){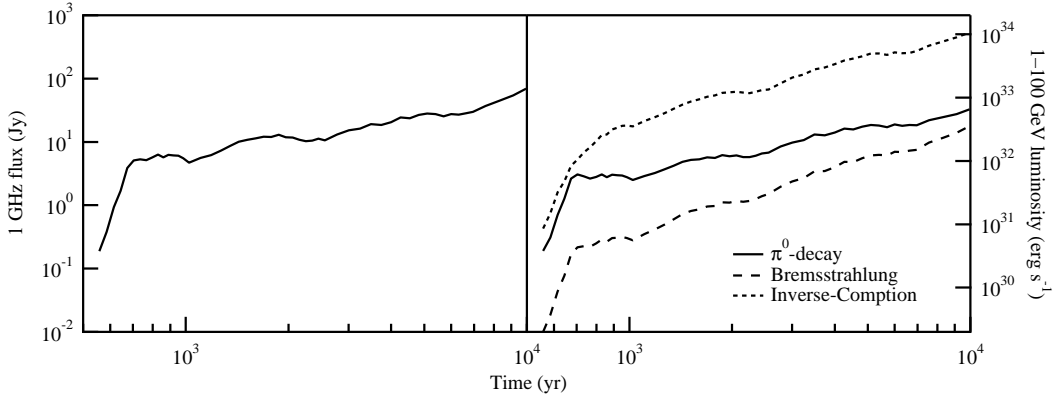}
  \end{center}
 \caption{{\it Left}: Radio flux at 1~GHz of synchrotron radiation from the blast-shocked ISM shell as functions of the elapsed time after explosion, calculated for $\xi_{\mathrm{e}}=10\,\xi_{\mathrm{p}}$ and $d=8$~kpc.  {\it Right}: Luminosities of bremsstrahlung (broken), inverse-Compton scattering (dotted) and $\pi^0$-decay (solid) $\gamma$-rays from the blast-shocked ISM shell in the 1--100~GeV band, as functions of the elapsed time after the explosion, calculated for $\xi_{\mathrm{e}}=10\,\xi_{\mathrm{p}}$.}  
  \label{fig4}
\end{figure*}

We calculate the non-thermal emission from the blast-shocked ISM shell, in which accelerated particles are confined for the either shorter time, $t_{\rm age}$ or $R_\mathrm{b}^2/D$.  Here $D$ is the diffusion coefficient and taken to be in the form $D_{\mathrm{10}}(E/10~\mathrm{GeV})(B/10~\mu\mathrm{G})^{-1}~\mathrm{cm}^2~\mathrm{s}^{-1}$ with a numerical coefficient $D_{\mathrm{10}}$.
Observations of cosmic-rays suggest that $D \sim10^{28}$~cm$^2$~s$^{-1}$ at 10~GeV \citep{Berezinskii90}. On the other hand, near SNRs, GeV and TeV observations suggest $D \sim10^{26}$~cm$^2$~s$^{-1}$ at 10~GeV (e.g., \cite{Li12}). We adopt $D_{10}=3\times10^{27}$ so that $D \sim 10^{28}$~cm$^2$~s$^{-1}$ for $B=3~\mu$G, typical field in the interstellar space, and $D \sim10^{26}$~cm$^2$~s$^{-1}$ for $B\sim100~\mu$G, which could be attained for SNRs. 

\subsection{Synchrotron radio}
Synchrotron radiation at the frequency $\nu=1$~GHz is mainly emitted by electrons with energies $\sim2\,(\nu/1\mathrm{~GHz})^{1/2}(B/100~\mu\mathrm{G})^{-1/2}$~GeV. According to \citet{Ginzburg65}, a flux $F_{\mathrm{syn}}(\nu)$ of synchrotron radiation emitted from electrons with the broken power-law spectrum is 
\begin{eqnarray}
&&F_{\mathrm{syn}}(\nu)\nonumber\\
&&=\frac{1}{4\pi d^2}\int\frac{\sqrt{3}e^3B^{(\mu+1)/2}K_{\mathrm{e}}}{2m_{\mathrm{e}}c^2}\left(\frac{16m_{\mathrm{e}}^3c^5\nu}{3e}\right)^{-(\mu-1)/2}\times\nonumber\\
&&\left[\int_{x_{\mathrm{b}}}^{x_{\mathrm{inj}}} F(x)x^{(\mu-3)/2}dx+E_{\mathrm{b}}^{-\mu+\mu_2}\left(\frac{16m_{\mathrm{e}}^3c^5\nu}{3eB}\right)^{(\mu-\mu_2)/2}\right.\nonumber\\
&&\left.\times\int_{x_{\mathrm{max}}}^{x_{\mathrm{b}}} F(x)x^{(\mu_2-3)/2}dx\right]4\pi r^2dr,
\end{eqnarray}
where
\[x=\frac{16m_{\mathrm{e}}^3c^5\nu}{3eBE^2},\]
$K_{\mathrm{e}}$ is $K$ of electrons (see Eq. (3)), $d$ is the distance to the SNR, $r$ is the radius from the center of the SNR, and $F(x)$ is the synchrotron function. The characters ``inj", ``b", and ``max" are correspond to the injection, break, and maximum energy, respectively. Integration interval of $r$ is given by the shocked ISM shell defined in the beginning of Section 3. Figure \ref{fig4} shows the time evolution of the 1~GHz flux for the magnetic field in Figure \ref{fig3}. 
Since the blast wave is little decelerated, the radio flux continues to increase through $\sim 10000$~yr (see Section 4). 

\subsection{Bremsstrahlung $\gamma$-ray}
Relativistic electrons emit bremsstrahlung $\gamma$-rays by interacting with target protons. Number of the bremsstrahlung photons emitted from electrons with the broken power-law spectrum per unit time per unit energy per unit volume is \citep{Blumenthal70}
\begin{eqnarray}
&&\frac{dN_{\gamma}}{dtdh\nu dV}
\nonumber\\
&&=\frac{4\alpha r_0^2cn_{\mathrm{T}}}{h\nu}\left[\ln\left(\frac{4h\nu}{m_{\mathrm{e}}c^2}\right)-\frac{1}{2}\right]\nonumber\\
&&\times \int^{E_{\mathrm{max}}}_{h\nu}dEN_{\mathrm{e}}E^{-2}\left[\frac{4}{3}E^2-\frac{4}{3}Eh\nu+(h\nu)^2\right]\nonumber\\
&&\simeq4\alpha r_0^2cK_{\mathrm{e}}n_{\mathrm{T}}\left[\ln\left(\frac{4h\nu}{m_{\mathrm{e}}c^2}\right)-\frac{1}{2}\right]\nonumber\\
&&\times\left\{
\begin{array}{ll}
\frac{(3\mu^2+\mu+4)(h\nu)^{-\mu}}{3\mu(\mu-1)(\mu+1)}-\frac{4(\mu_2-\mu)E_{\mathrm{b}}^{-\mu+1}(h\nu)^{-1}}{3(\mu-1)(\mu_2-1)}\\+\frac{4(\mu_2-\mu)E_{\mathrm{b}}^{-\mu}}{\mu\mu_2}-\frac{(\mu_2-\mu)E_{\mathrm{b}}^{-\mu-1}h\nu}{(\mu+1)(\mu_2+1)},~h\nu \leq \frac{1}{2}E_{\mathrm{b}},\\
\frac{(3\mu_2^2+\mu_2+4)(h\nu)^{-\mu_2}}{3\mu_2(\mu_2-1)(\mu_2+1)},~h\nu > \frac{1}{2}E_{\mathrm{b}},\\
\end{array}
\right.
\label{equation11}
\end{eqnarray}
where $\alpha$ is the fine structure constant, $r_0$ is the classical electron radius, $h$ is the Planck constant, and $n_{\mathrm{T}}$ is the number density of target protons. The target protons are assumed to be thermal protons in the shocked ISM shell. We consider $E_{\mathrm{max}}\gg h\nu$ in the last expression in equation (\ref{equation11}). Using equation (\ref{equation11}), we get a bremsstrahlung $\gamma$-ray luminosity by
\begin{equation}
L_{\mathrm{brems}}=\int \int h\nu\frac{dN_{\gamma}}{dtdh\nu dV}4\pi r^2drdh\nu. 
\end{equation}
Integration interval of $r$ is given by the shocked ISM shell defined in the beginning of Section 3. We calculate the luminosity using $\langle n^2 \rangle$ obtained from hydrodynamical calculation of Paper I. Figure \ref{fig4} shows the time evolution of $\gamma$-ray luminosity due to bremsstrahlung in the 1--100 GeV band.

\subsection{Inverse-Compton $\gamma$-ray}
When a photon of energy $\epsilon_0$ is scattered by a relativistic electron of Lorentz factor $\gamma$, scattered photon has energy of $\sim\gamma^2\epsilon_0$. Because the maximum energy of accelerated electrons is $\sim10$~TeV, the electrons can emit photons of energies up to $\sim$100~GeV by scattering cosmic microwave background (CMB) photons. Number of photons emitted from electrons with the broken power-law spectrum by scattering CMB photons of the temperature $T_{\mathrm{CMB}}$ per unit time per unit energy per unit volume is \citep{Blumenthal70}
\begin{eqnarray}
&&\frac{dN_{\gamma}}{dtdh\nu dV}\nonumber\\
&&=\frac{8\pi^2 r_{0}^2K_{\mathrm{e}}E_{\mathrm{b}}^{-\mu+\mu_2}}{h^3c^2(m_\mathrm{e}c^2)^{\mu_2-1}}(kT_{\mathrm{CMB}})^{(\mu_2+5)/2}\left[(h\nu)^{-(\mu_2+1)/2}\times \right.\nonumber\\
&&\frac{2^{\mu_2+3}(\mu_2^2+4\mu_2+11)}{(\mu_2+3)^2(\mu_2+5)(\mu_2+1)}\Gamma\left(\frac{\mu_2+5}{2}\right)\zeta\left(\frac{\mu_2+5}{2}\right)\nonumber\\
&&\left.-h\nu\frac{2^{\mu_2+2}\pi^2}{3(\mu_2+1)}\left(\frac{m_{\mathrm{e}}^2c^4}{4E_{\mathrm{max,e}}^2kT_{\mathrm{CMB}}}\right)^{(\mu_2+1)/2}\right],
\label{equation13}
\end{eqnarray}
where $\zeta$ is the zeta function and $E_{\mathrm{max,e}}$ is the maximum energy of accelerated electrons. Using equation (\ref{equation13}), we get a inverse-Compton $\gamma$-ray luminosity by
\begin{equation}
L_{\mathrm{IC}}=\int\int h\nu\frac{dN_{\gamma}}{dtdh\nu dV}4\pi r^2drdh\nu.
\end{equation}
Integration interval of $r$ is given by the shocked ISM shell defined in the beginning of Section 3. Figure \ref{fig4} shows the time evolution of the $\gamma$-ray luminosity due to inverse-Compton scattering in the 1--100~GeV band.

\subsection{$\pi^0$-decay $\gamma$-ray}
Relativistic protons emit neutral $\pi^0$s through inelastic collisions with protons, and then the $\pi^0$s decays into two $\gamma$-ray photons. We calculate the $\pi^0$-decay $\gamma$-ray luminosity, using the parameterized cross section of inelastic proton-proton collision
\begin{equation}
\sigma_{\mathrm{inel}}(\tilde{E}_{\mathrm{p}})\simeq3\left[0.95+0.06\ln\left(\frac{E_{\mathrm{p}}}{1~\mathrm{GeV}}\right)\right]\times10^{-26}\mathrm{~cm}^2,
\end{equation}
and $\delta$-function approximation of number of $\pi^0$s emitted per unit time per unit energy per unit volume
\begin{eqnarray}
&&\frac{dN_{\pi}}{dtd\tilde{E}_{\pi}dV}\nonumber\\
&&=\frac{cn_{\mathrm{T}}}{f_{\pi}}\sigma_{\mathrm{inel}}\left(m_{\mathrm{p}}c^2+\frac{\tilde{E}_{\pi}}{f_{\pi}}\right)N_{\mathrm{p}}\left(m_{\mathrm{p}}c^2+\frac{\tilde{E}_{\pi}}{f_{\pi}}\right),
\end{eqnarray}
which are used in \citet{Aharonian00}. Here $f_{\pi}\simeq0.17$ is mean fraction of the kinetic energy of proton transferred to $\pi^0$ per collision, $\tilde{E}_{\mathrm{p}}=\gamma m_{\mathrm{p}}c^2$ is total energy of proton and $\tilde{E}_{\pi}=\gamma m_{\pi}c^2$ is total energy of $\pi^0$. Number of $\pi^0$-decay photons emitted from protons with the broken power-law spectrum per unit time per unit energy per unit volume is
\begin{eqnarray}
&&\frac{dN_{\gamma}}{dtdh\nu dV}\nonumber\\
&&=2\int^{\infty}_{E_{\mathrm{min}}}\frac{1}{(\tilde{E}_{\pi}^2-m_{\pi}^2c^4)^{1/2}}\frac{dN_{\pi}}{dtd\tilde{E}_{\pi}dV}d\tilde{E}_{\pi}\nonumber\\
&&\simeq2\int^{\infty}_{h\nu}\frac{1}{\tilde{E}_{\pi}}\frac{dN_{\pi}}{dtd\tilde{E}_{\pi}dV}d\tilde{E}_{\pi}\nonumber\\
&&\simeq3\times10^{-26}\frac{2cn_{\mathrm{T}}K_{\mathrm{p}}}{f_{\pi}}\times\nonumber\\
&&\left\{
\begin{array}{ll}
\frac{1}{\mu}\left(\frac{h\nu}{f_{\pi}}\right)^{-\mu}\left[0.95+0.06\left(\ln\left(\frac{h\nu/f_{\pi}}{1\mathrm{~GeV}}\right)+\frac{1}{\mu}\right)\right]\\
+E_\mathrm{b}^{-\mu}\left[0.95\left(\frac{1}{\mu_2}-\frac{1}{\mu}\right)+0.06\left(\frac{1}{\mu_2^2}-\frac{1}{\mu^2}\right)\right.\\
\left.+0.06\left(\frac{1}{\mu_2}-\frac{1}{\mu}\right)\ln\left(\frac{E_\mathrm{b}}{1~\mathrm{GeV}}\right)\right],\\
\mathrm{for}~h\nu\leq f_{\pi}E_\mathrm{b},\\
\frac{E_\mathrm{b}^{-\mu+\mu_2}}{\mu_2}\left(\frac{h\nu}{f_{\pi}}\right)^{-\mu_2}\left[0.95+0.06\left(\ln\left(\frac{h\nu/f_{\pi}}{1\mathrm{~GeV}}\right)+\frac{1}{\mu_2}\right)\right],\\
\mathrm{for}~h\nu> f_{\pi}E_\mathrm{b},
\end{array}
\right.
\label{equation17}
\end{eqnarray}
where $E_{\mathrm{min}}=h\nu+(m_{\pi}^2c^4/4h\nu)$ is the minimum pion energy to produce photon of energy $h\nu$ and $K_{\mathrm{p}}$ is $K$ of protons (see Eq. (3)).  In the second expression in equation (\ref{equation17}), $h\nu>m_{\pi}c^2$ and $\tilde{E}_\pi>m_{\pi}c^2$ are considered because we calculate photons above $1~$GeV. In the last expression of equation (\ref{equation17}), we approximate the proton energy spectrum as the relativistic form and the variable of the spectrum as $(m_{\mathrm{p}}c^2+\tilde{E}_{\pi}/f_{\pi})\sim \tilde{E}_{\pi}/f_{\pi}$, because $\tilde{E}_{\pi}/f_{\pi}\gtrsim (1\mathrm{~GeV}/0.17)\sim6\mathrm{~GeV}>m_{\mathrm{p}}c^2$. Using equation (\ref{equation17}), we get a $\pi^0$-decay $\gamma$-ray luminosity by
\begin{equation}
L_{\pi}=\int \int h\nu\frac{dN_{\gamma}}{dtdh\nu dV}4\pi r^2drdh\nu.
\end{equation}
Integration interval of $r$ is given by the shocked ISM shell defined in the beginning of Section 3. As in the calculation of the bremsstrahlung luminosity, we use $\langle n^2 \rangle$ instead of $\langle n\rangle^2$. Figure \ref{fig4} shows the time evolution of $\gamma$-ray luminosity due to $\pi^0$-decay in the 1--100 GeV band.

\section{Discussion}

For the low density ISM of density $0.016~$cm$^{-3}$ in the present model, supposed for an H\,II region (e.g., formed by the progenitor and extended to a few tens pc), the blast wave is little decelerated through $\sim 10000$~yr.  As a result, in the context of diffusive shock acceleration described in Section 2, the radio flux continues to increase, because the increase of the emission measure overcomes the decrease of the magnetic field strength.  
Consequently, for about ten thousand years, recombination-radiation X-rays are observed from the irregular-shape inner part of SNR (see Figure \ref{fig1} and Paper I), while the radio emission of tens Jy is observed from the blast-shocked ISM shell. 

In the beginning of the Sedov/Taylor phase where the blast wave is being decelerated significantly as $V_{\rm s} \propto t^{-3/5}$, the radio flux turns to decrease slowly as $t^{-3/10}$, and then approaches nearly constant as the magnetic field approaches its interstellar value ($\sim3$~$\mu$G) and $T_{\rm e}$ approaches $T_{\rm p}$.  
Also the inverse-Compton $\gamma$-rays turns to decrease as $\propto t^{-1/5}$ in the Sedov/Taylor phase, while $\pi^0$-decay $\gamma$-rays are nearly constant.  This sort of analysis is done also for the phase $\lesssim 10000$~yr with the relation $V_{\rm s} \propto t^{-s}$ where $s$ is given by the hydrodynamical calculation, and gives a good agreement with the computed time evolution of the radio and $\gamma$-ray emission in Figure \ref{fig4}.  It should be noted that $s \sim 0.4$ at 10000~yr, yet smaller than the Sedov value $s = 3/5$, and the SNR is in the transient phase to the Sedov/Taylor regime. 

Again because of low density, the $\gamma$-ray luminosity of the shocked ISM shell is dominated by inverse-Compton scattering through the SNR evolution concerned. 
However, $\pi^0$-decay $\gamma$-rays could be enhanced by interactions with dense external matter, e.g., dense H\,I gas, molecular clouds or a cavity wall formed by the stellar wind of the progenitor. If $10$\% of accelerated protons interact with such matter of density $n\sim100~$ cm$^{-3}$, the luminosity $L_{\pi}$ would exceed $10^{35}$~erg~s$^{-1}$ at a few thousands year, comparable to the typical $\gamma$-ray luminosity of MM SNRs. The interactions with molecular clouds are suggested in many MM SNRs by OH maser and/or near-infrared observations. The interaction with HI gas is suggested in RX J1713.7-3946 by observations \citep{Fukui12}, and may be expected also in MM SNRs. 

\begin{figure}[htpb]
  \begin{center}
    \FigureFile(83mm,88mm){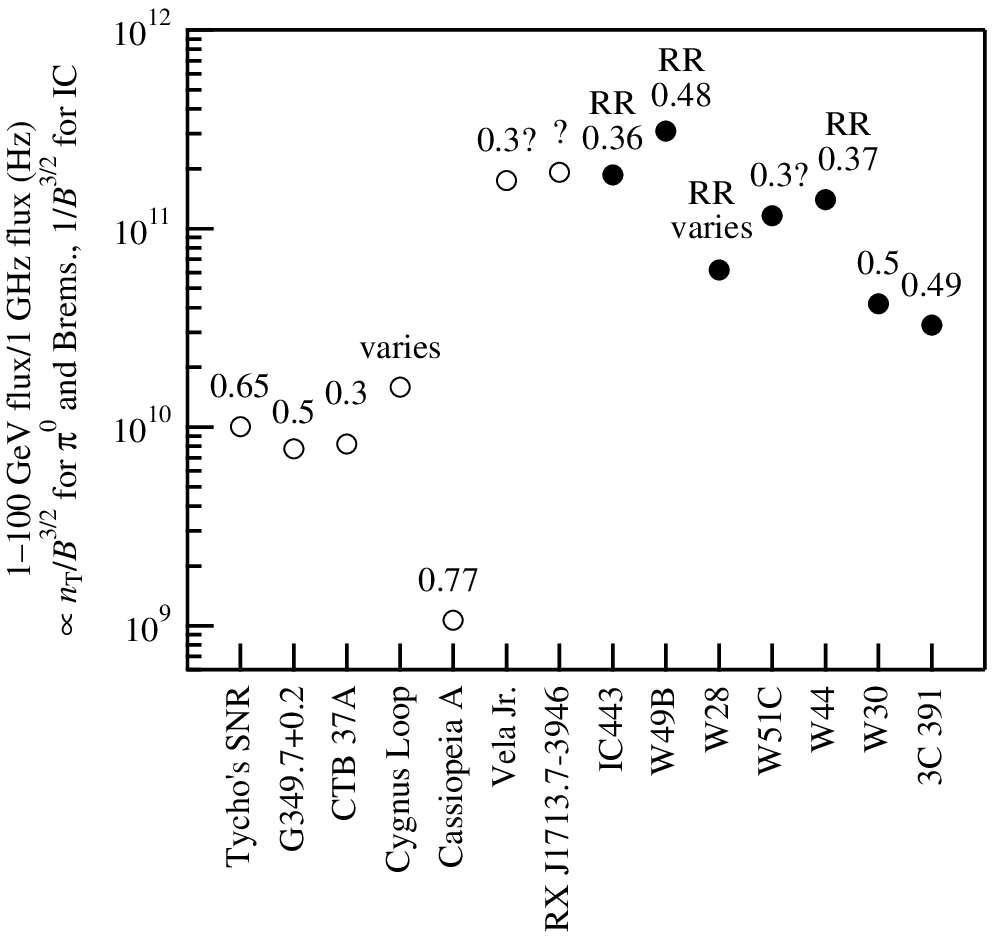}
  \end{center}
  \caption{Flux ratio of 1--100~GeV to 1~GHz of shell-like (open circles) and MM (filled circles) SNRs observed so far by {\it Fermi}. Attached numbers indicate the power indexes of the radio spectra \citep{Green09} and ``RR" means that recombination-radiation X-rays are observed.
\newline
References.---Radio flux of SNRs, except for RX J1713.7-3946 \citep{Acero09}; \cite{Green09}, $\gamma$-rays and distance:
Tycho's SNR; \cite{Hayato10}, \cite{Giordano11},
G349.7+0.2; \cite{Frail96}, \cite{Castro10},
CTB 37A; \cite{Reynoso00}, \cite{Castro10},
Cygnus Loop; \cite{Blair05}, \cite{Katagiri11},
Cassiopeia A; \cite{Reed95}, \cite{Abdo10b},
Vela Jr.; \cite{Katsuda08}, \cite{Tanaka11},
RX J1713.7-3946; \cite{Fukui03}, \cite{Abdo11}, 
IC443; \cite{Welsh03}, \cite{Abdo10c},
W49B; \cite{Moffett94}, \cite{Abdo10e},
W28; \cite{Vel02}, \cite{Abdo10d},
W51C; \cite{Koo95}, \cite{Abdo09},
W44; \cite{Wols91}, \cite{Abdo10a},
W30; \cite{Fich89}, \cite{Castro10},
3C 391; \cite{Frail96}, \cite{Castro10}
  }
    \label{fig5}
\end{figure}

Interaction with some dense external matter may be realized on the $\gamma$-ray to radio flux ratio. 
We show the ratio of 1--100 GeV to 1 GHz flux in Figure \ref{fig5}. 
One can see that the ratio is systematically high for MM SNRs compared to shell-like SNRs except for RX J1713.7-3946 and Vela Jr. 
For RX J1713.7-3946 interaction with molecular clouds (e.g., \cite{Dame01}) and/or H\,I gas (\cite{Fukui12}) is suggested by observations. 
As for Vela Jr., an X-ray source CXOU J085201.4-461753, possible neutron star, is located near the center of the SNR \citep{Pavlov01}.  That might be related to the hard radio spectrum and high $\gamma$-ray/radio ratio, though the pulsar activity is not observed. 
The low ratio of Cas A is due likely to a strong field $\sim1$~mG (e.g., \cite{Arbutina12}).


When electrons responsible for radio and electrons/protons for $\gamma$-rays are produced in the same volume, the $\gamma$-ray/radio ratio is reduced to $n_{\mathrm{T}}/B^{3/2}$. 
Therefore, if the particle spectrum and $B$ do not vary so much from SNR to SNR, the ratio can be a measure of the density $n_{\mathrm{T}}$ of the matter with which the particles interact. 
For $B \sim 100$~$\mu$G the high ratios observed from MM SNRs may be explained by $\pi^0$-decay if the density of the target matter $n_{\mathrm{T}} > 10~$cm$^{-3}$, higher than the typical ISM density ($\lesssim 1$~cm$^{-3}$). 
The high ratios could be explained also by inverse-Compton if $B\lesssim10~\mu$G. 
Such a field may be possible for shell-like SNRs, but unlikely for MM SNRs which exhibit rather high radio flux. 

Finally, we mention the effect of the CSM, stellar wind matter here. An important effect of the CSM is that the shock break-out raises the maximum energy $E_{\mathrm{max}}$ to $\sim1300$~TeV for protons (see Figure \ref{fig3}). Since $E_{\mathrm{max}}\propto BV_{\mathrm{s}}^2t\propto V_{\mathrm{s}}^3 \propto (E_{\mathrm{ej}}/M_{\mathrm{ej}})^{3/2}$, where $E_{\mathrm{ej}}$ and $M_{\mathrm{ej}}$ are the initial kinetic energy of ejecta and the ejecta mass, respectively, $E_{\mathrm{max}}$ would reach $\sim3000$~TeV, the cosmic-ray knee energy, for 2 times larger value of  $E_{\mathrm{ej}}/M_{\mathrm{ej}}$ than that in the present model.

\bigskip
We are grateful to Yutaka Ohira, Inoue Tsuyoshi and Ryo Yamazaki for meaningful discussion about particle acceleration. Also to the anonymous referee for his/her careful reading the manuscript.  
KM and KK are respectively supported by the Grant-in-Aid for
Scientific Research 22540253 and 24540229, from Japan Society for the Promotion of Science (JSPS).

\end{document}